\documentclass[onecolumn,authoryear]{els-mrw} 

\usepackage{amsmath,amssymb,amsfonts,amsthm,makeidx,graphicx}
\usepackage{txfonts}
\usepackage{helvet}
\usepackage{soul}

\begin{document}

\chapter{The First Particles}\label{chap1}

\author[1]{Fa Peng Huang}

\address[1]{\orgname{Sun Yat-sen University (Zhuhai Campus)}, \orgdiv{MOE Key Laboratory of TianQin Mission, TianQin Research Center for
		Gravitational Physics \& School of Physics and Astronomy, Frontiers
		Science Center for TianQin, Gravitational Wave Research Center of
		CNSA}, \orgaddress{Zhuhai 519082, China}}

\articletag{Chapter Article tagline: update of previous edition,, reprint..}

\maketitle


\begin{glossary}[Nomenclature]
\begin{tabular}{@{}lp{34pc}@{}}
SM &Standard Model\\
BBN & Big Bang Nucleosynthesis\\
CMB &  Cosmic Microwave Background \\
\end{tabular}
\end{glossary}

\begin{abstract}[Abstract]
After cosmic inflation, the universe is cold and almost empty. Thus, the inflation field should decay to the particles for BBN through the so-called reheating process.
Later, the matter-antimatter asymmetry and dark matter are produced. In this chapter, the ``first particle" production between the inflation phase and BBN phase is introduced. We focus on the reheating, electroweak baryogenesis, and leptogenesis.
\end{abstract}

Key points
\begin{itemize}
\item{Reheating is the process in the early universe where the energy from cosmic inflation is transferred to particles, leading to the thermalization and initiation of the hot Big Bang.}
\item{Baryogenesis the theoretical mechanism that explains the observed matter-antimatter asymmetry of our universe, resulting in the predominance of matter.}
\item{Electroweak phase transition is the dynamic process during which the electroweak symmetry is spontaneously broken, a transition that occurs as the Higgs field acquires a nonzero vacuum expectation value.  The  phase transition dynamics are essential for explaining particle production in the early universe.}
\item{Leptogenesis is a theoretical process in the early universe that generates an imbalance between leptons and antileptons, which can later be converted into the observed matter-antimatter asymmetry.}

\end{itemize}

\begin{figure}[t]
\centering
\includegraphics[width=1.0\textwidth]{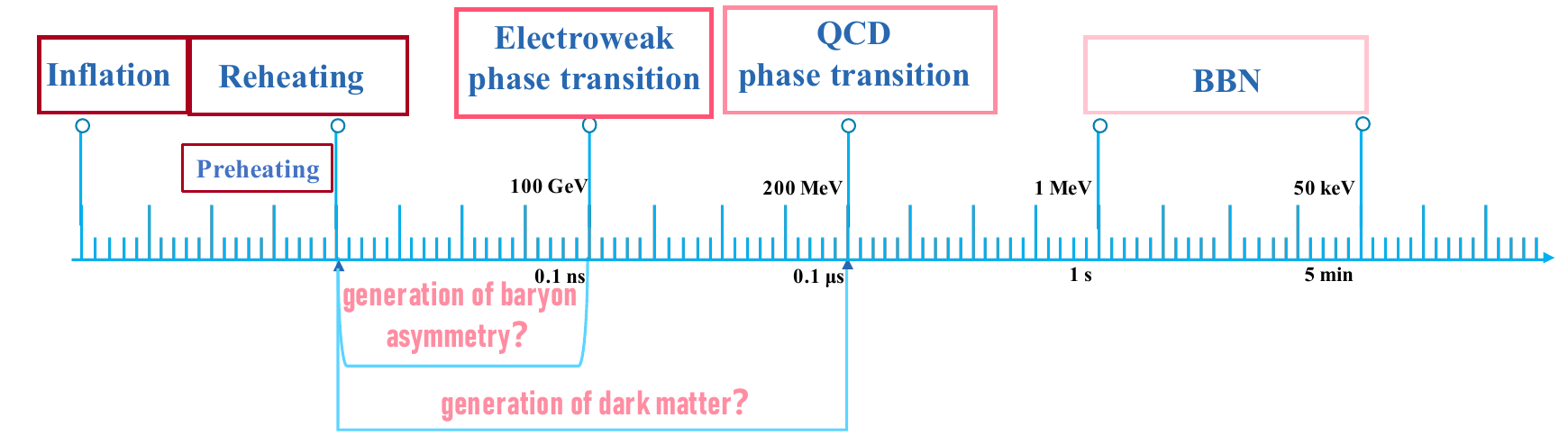}
\caption{The schematic process of the first particle production in the early universe. During the reheating process, the first particles are created from the late time of cosmic inflation and provide the source of BBN.}
\label{fig:evolve}
\end{figure}

\section{introduction}\label{firstparticles:intro}

In recent decades, the precise measurements of the Cosmic Microwave Background (CMB) and numerous other astronomical observations have profoundly changed our understanding of the universe. Modern cosmology, rooted in general relativity and particle physics, primarily relies on large-scale astronomical observations to study the fundamental properties of the universe and its evolution laws. The standard model (SM) of cosmology, $\Lambda$ cold dark matter ($\Lambda$CDM) model is established, where the current universe is composed of matter ($4.9\%$), dark matter ($26.8\%$), and dark energy ($68.3\%$) (\cite{Planck:2018vyg}). The origin and properties of these components are the central problems in astrophysics, cosmology, and particle physics. The two theoretical cornerstones of modern cosmology are the theories of cosmic inflation and Big Bang nucleosynthesis (BBN).
Cosmic inflation provides the primordial seeds of the large scale structure of our universe and BBN describes how the light elements in our universe are produced. There is an unknown but important period between the inflation and BBN.
During this period, the first particles, which are the particle seeds of BBN, should be produced. More precisely, the net baryon and dark matter should be produced in this period.
However, how are the particles produced after inflation?
How is the matter produced? How is the dark matter produced?

Inflationary cosmology is the standard paradigm of early universe physics. Inflation leaves the universe in a non-thermal state, cold, and effectively empty of matter. Hence, one must explain how the inflationary phase is connected to the high-temperature phase at BBN, as well as explain the production of the matter and dark matter. 
To achieve this, the reheating mechanism has been introduced to bridge the gap between inflation and BBN.
The inflaton or inflation field has to couple to the SM fields. The energy stored in the inflation field should be transferred to particles and create high-energy cosmic plasma through the reheating process. The universe’s temperature decreases as it expands, with its thermal evolution history illustrated in Fig. \ref{fig:evolve}. 
During the high-energy reheating phase, particles of the early universe, potentially including dark matter, may be generated. Around 
100 GeV, the universe undergoes the electroweak phase transition, a period during which baryon asymmetry might be produced through electroweak baryogenesis. At approximately 
0.1 MeV, the universe enters BBN stage, leading to the formation of the lightest nuclides.

In this chapter, we introduce particle production after inflation and focus on the generation of matter-antimatter asymmetry through electroweak baryogenesis and leptogenesis. 
See other chapters for a detailed introduction to inflation, BBN, and dark matter including potential production mechanisms.

\section{Reheating and preheating after inflation}\label{firstparticles:sec1}

\subsection{Why do we need reheating?}

Our understanding of the early evolution of the universe mainly relies on inflation and BBN. Both of them are built on solid theoretical foundations and are supported by high-precision measurement data. According to modern cosmological theories, the universe at the end of inflation is in a low-temperature state. The energy of the inflatons, which drive the inflation, must be transferred to particles, eventually resulting in a high-temperature, thermodynamically balanced, radiation-dominated universe that sets the stage for BBN. The transition of the universe from the supercooled state at the late stage of inflation to the overheated, radiation-dominated state before BBN is known as reheating. In the reheating period, the fundamental particles of the SM emerged in the early universe (\cite{Kofman:1994rk, Kofman:1997yn, Lozanov:2020zmy}). The first stage of the reheating process, namely, the particle production process, is called preheating.
Reheating plays a crucial role in explaining the origin of key components of the universe, such as the baryonic matter, photons, neutrino background observable, the potential production of dark matter, and gravitational waves.

\subsection{Reheating dynamics }

\begin{figure}
    \centering
    \includegraphics[width=0.5\linewidth]{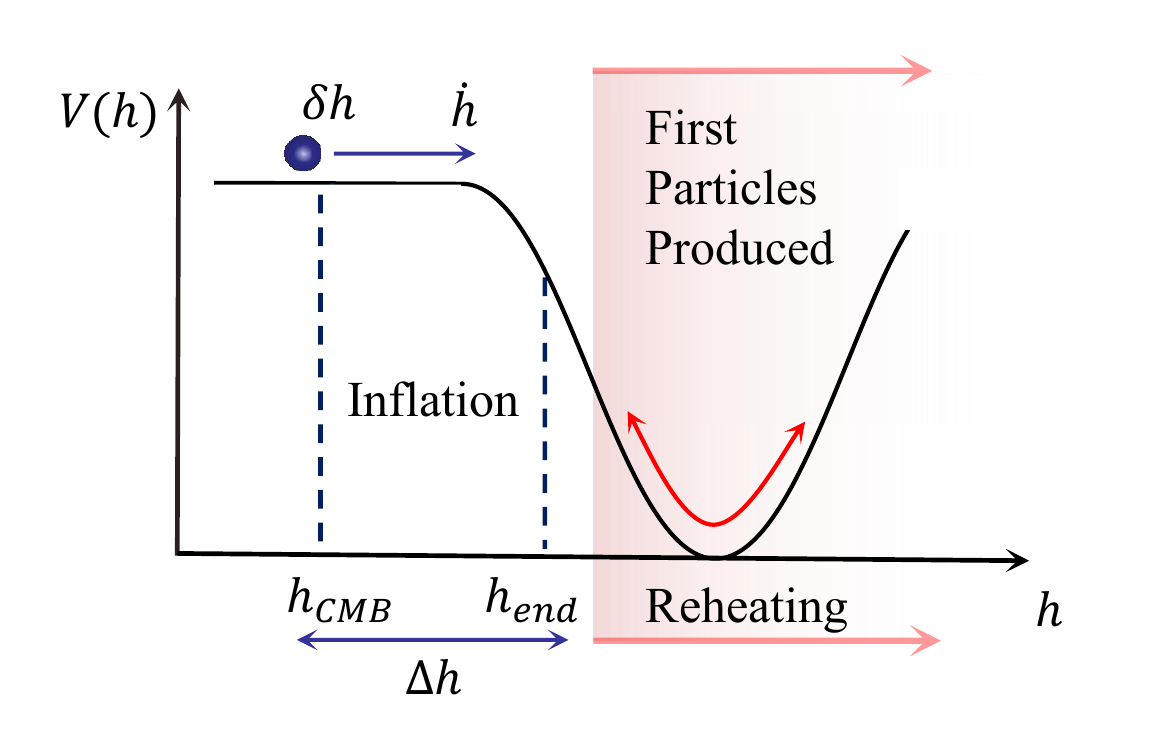}
    \caption{The reheating process from the end of cosmic inflation. The coherent oscillation of the inflation field plays an essential role in the scenario. The horizontal axis represents the inflation field $h$ value and 
the vertical axis represents the potential energy density
$V(h)$ of the inflation field in the early universe. $h_{CMB}$ is the field value at the early stage of inflation which can be tested by the CMB observation. $h_{end}$ is the field value at the end of inflation.
$\delta h$ represents the quantum fluctuations of the inflation field $h$ and $\dot{h}$ means the time derivative of the inflation field.}
    \label{fig:enter-label}
\end{figure}

As shown in Fig. \ref{fig:enter-label},
at the late stage of cosmic inflation, the early universe is believed to have evolved in a specific manner. Regardless of the underlying dynamics, the energy density of the inflaton will be eventually converted into radiation energy density. This process is known as reheating.
Reheating marks the transition from the inflationary phase to a hot, radiation-dominated era. As the inflaton field oscillates around the minimum of its potential, it transfers its energy to other fields, generating a plethora of particles as shown in Fig. \ref{fig:enter-label}. The energy transfer process is highly non-linear and can produce particles with a wide range of energies. The reheating process can be roughly divided into two stages.

Stage 1 (preheating from resonance): In the first stage, the oscillation amplitude of the inflation field is large and coherent. The inflaton has to couple to the SM fields to avoid the universe becoming completely empty. The energy stored in the inflation field will then be transferred to ordinary particles. If the inflaton only couples to fermions, the decay of the inflaton is usually slow. If the inflaton couples to bosons, then the decay rate may be greatly increased due to Bose condensation and parametric resonance effects. 
Similar to the effect of Bose condensation, the energy conversion efficiency is significantly enhanced in the presence of multiple particles. Due to the presence of resonance effects (including self-resonance, parametric resonance, and tachyonic resonance), the energy conversion efficiency is higher. The parametric resonance and tachyonic resonance describe the process of transferring energy from the inflaton condensate to the fluctuations of the decay product particles. Self-resonance involves the process of transferring energy from the inflaton condensate to the fluctuations of the inflaton field. Anyone who has been on a swing understands this concept of parametric resonance. By pumping your legs at the right moment, you transfer energy to the swing, making it go higher.
The parametric resonance describes how the inflaton field's oscillations after inflation amplify fluctuations in other fields it interacts with. These oscillations cause periodic variations in the effective mass of these fields, creating resonance that leads to rapid, exponential particle production.
The tachyonic resonance is a process where fields coupled to the oscillating inflaton experience negative effective mass-squared values, making them behave like "tachyons" (hypothetical particles with imaginary mass). This causes rapid, exponential growth of fluctuations in these fields, leading to efficient particle production. 
This first phase with rapid creation of particles is known as the preheating phase since the particles are generated far from thermal equilibrium.
The preheating phase is usually associated with resonances and leads to the explosive production of new particles. 
 
Stage 2 (perturbative reheating): Subsequently, the early universe enters into the perturbative heating phase. Later, the produced particles scatter off each other and come to some sort of thermal equilibrium, reminiscent of BBN conditions. 

In the high-energy environment of reheating, early universe particles, including dark matter, may be produced. The first particles formed during this process are often high-energy bosons, such as gauge bosons (like photons and gluons) and potentially scalar particles (if the inflaton field interacts with scalar fields). These particles undergo rapid thermalization, creating a hot and dense plasma. As the universe expands and cools, more particles are generated through various interactions, eventually leading to the production of familiar particles observed in the present-day universe, such as electrons, quarks, and neutrinos. Reheating thus sets the stage for the subsequent evolution of the universe, ultimately leading to the formation of galaxies, stars, and planets.

In reality, the dynamics during reheating can be very complicated, often involving non-perturbative effects that must be calculated by numerical simulations. Describing a special but important period in the early universe is essential for understanding how the hot BBN began and the study of reheating remains a very active area of research. 

\subsection{Implication and detection of reheating}

At the end of cosmic inflation, reheating is essential for producing the first particles for BBN and the late universe. In many cases, the reheating process can also generate various relics such as the matter-antimatter asymmetry of the universe (see next section for details of this asymmetry),  dark matter or radiation, isocurvature perturbations, stochastic gravitational waves, non-Gaussianities, primordial black holes, topological and non-topological solitons, and primordial magnetic fields~(\cite{Kofman:1994rk, Kofman:1997yn, Lozanov:2020zmy}).  
Detecting any of the reheating relics would provide a fascinating window into the reheating era. Especially, various studies have shown that the reheating process could produce high-frequency gravitational wave signals.
Among these possible relics, the stochastic gravitational waves 
might provide a unique way to explore the underlying physics of reheating.

\section{Matter-antimatter asymmetry and baryogenesis}\label{firstparticles:sec2}
The origin of the matter-antimatter asymmetry of the universe is one of the central issues of particle cosmology, and has been a long unsolved problem in cosmology, astrophysics, and particle physics. It is an important part of the puzzle of the energy budget of the universe.

Although it is challenging to solve this problem, our understanding of it becomes deeper with the improvement of theoretical and experimental studies. Firstly, we introduce the experimental background of the matter-antimatter asymmetry of the universe and show the three necessary conditions to solve the problem. Then, we review the current status of our understanding of this problem, emphasizing scenarios that can be tested in the experiments, including electroweak baryogenesis and leptogenesis. We hope the experiments in the near future can unravel the true baryogenesis scenario, which can explain the matter-antimatter asymmetry of the universe.

\subsection{The observed matter-antimatter asymmetry of the universe}

The familiar ordinary matter we know is composed of elementary particles in the SM of particle physics. This part of the matter only constitutes about $4.9\%$
of the total energy in the universe~(\cite{Planck:2018vyg}). However, even this $4.9\%$ of ordinary matter presents a puzzling mystery of the universe, which is the problem of matter-antimatter asymmetry.

From theoretical prediction, it would be most natural for there to be equal amounts of matter and antimatter in the universe. In 1928, Dirac formulated the relativistic motion equation for microscopic particles, known as the Dirac equation, which first predicted the existence of antimatter. For decades, experimental particle physics has confirmed that each particle has a corresponding antiparticle (the antiparticle of a photon is itself). When matter meets antimatter, they annihilate into photons carrying corresponding energy. From the perspective of microscopic particle physics, matter and antimatter are equivalent, neither superior to the other.
However, applying this concept to the universe has caused great confusion. 

From experiments, however, observational evidence from astronomy and cosmology indicates a severe imbalance in their status in the universe. In the vast universe, people only observe matter and not antimatter. The asymmetry between matter and antimatter in the universe mainly manifests as an asymmetry between baryons and antibaryons because the energy of ordinary matter in today's universe is mainly concentrated in baryons. If there were regions of antimatter in the universe, strong gamma rays would be emitted at the boundaries of these regions due to matter-antimatter annihilation. However, to date, such gamma rays have not been observed (\cite{Cohen:1997ac}). 
Indeed, antimatter is usually detected in accelerators or cosmic rays. Antimatter observed in high-energy cosmic rays, such as antiprotons, is produced as secondary particles in collisions during the propagation of cosmic rays and does not originate from primordial antimatter in the depths of the universe. Detailed analysis results indicate no regions of antimatter within the observable universe. 
Essentially, the observable universe, out to the Hubble size, is made of matter and not antimatter (\cite{Cohen:1997ac}).

In cosmology, people often define the baryon-to-photon ratio $\eta_B = (n_B - \bar{n}_B) / n_{\gamma}$ to quantitatively describe the asymmetry between matter and antimatter in the universe (where $n_B$ and $\bar{n}_B$ represent the number densities of baryons and antibaryons, and $n_{\gamma}$ is the number density of photons, approximately 413 photons per cubic centimeter). Classical Big Bang cosmology tells us that in the early universe, matter was in a hot plasma state. When the universe's temperature was high enough, baryons and antibaryons were constantly created in pairs and quickly annihilated. However, when the temperature dropped below 1 GeV, these baryons and antibaryons quickly annihilated into photons, and no more pairs were produced. If the universe were symmetric in baryons and antibaryons, the final result would be $\eta_B = 0$. However, this is directly contradicted by observational results. 

As the observational evidence from cosmology, the BBN data give
\begin{align}\label{eq:bbn}
	\eta_{B}=(6.040 \pm 0.118)\times 10^{-10} \quad (\mathrm{BBN})
\end{align}
and CMB radiation data give
\begin{align}\label{eq:cmb}
	\eta_{B}=(6.12 \pm 0.048)\times 10^{-10}  \quad (\mathrm{CMB})
\end{align}
Both provide similarly precise measurements of the baryon-to-photon ratio, 
\begin{align}\label{eq:cmbbn}
	\eta_{B}=(6.115 \pm 0.038)\times 10^{-10} \quad (\mathrm{combined})
\end{align}
which corresponds to $\Omega_{\mathrm{b}} h^2=0.02233 \pm 0.00014$ (\cite{Yeh:2022heq}).
Note that these two physical processes occur at vastly different temperatures or energy scales, differing by a factor of a million, yet they yield almost identical results. This demonstrates the consistency and success of the standard Big Bang cosmological model, which consistently tells us that at least from the time of nucleosynthesis, the universe has exhibited a significant asymmetry between matter and antimatter.

Theoretically, this asymmetry between matter and antimatter may have existed since the birth of the universe. However, recent cosmological studies suggest that the universe underwent a period of rapid expansion in its early stages. Inflation solved problems such as the flatness and uniformity of the classical universe, but the violent expansion also caused the original densities of baryons and antibaryons to approach zero. In other words, after inflation, the universe should be in a symmetric state with $\eta_B = 0$. All matter and antimatter were produced during the reheating process after inflation, so the asymmetry between matter and antimatter must also result from the universe's dynamical evolution after inflation.

So, how was this asymmetry between matter and antimatter created?
In other words, what kind of physical processes would cause the universe to evolve from a state of $\eta_B = 0$ to an asymmetrical state with $\eta_B=6 \times 10^{-10}$?

\subsection{Sakharov conditions and baryogenesis}

As early as 1967, Sakharov proposed the famous three conditions for the dynamical generation of the asymmetry between matter and antimatter in the universe, known as the baryogenesis mechanism (\cite{sakharov1991pis}).
\begin{itemize}%
\item[$\circ$] Baryon number (B) violation:
 
 The first condition requires the existence of a physical process where the baryon number is not conserved. This is quite evident; if the baryon number is conserved, the universe with matter-antimatter symmetry would always remain symmetrical.
 
\item[$\circ$] Charge (C) and Charge-Parity (CP) violation:
 
The second condition is the violation of C and CP symmetries. C asymmetry refers to the asymmetry in the exchange of particles and antiparticles, while CP asymmetry is the combined asymmetry in the exchange of particles and antiparticles, as well as left- and right-handed exchanges. As long as either C or CP symmetry exists, reaction processes that break the baryon number will produce an equal number of baryons and antibaryons. 

\item[$\circ$] Departure from thermal equilibrium or Charge-Parity-Time (CPT) violation:
 
The third condition is departure from thermal equilibrium. According to the CPT theorem (where T represents time reversal), particles and antiparticles have equal masses. If the system were in thermal equilibrium, baryons and antibaryons would have the same thermal distribution, resulting in equal densities and quantities. 
\end{itemize}

Since the 1970s, various mechanisms fulfilling Sakharov's three conditions have been proposed by physicists (\cite{Dine:2003ax}). Typical examples include grand unified theory baryogenesis, the Affleck-Dine mechanism, gravitational baryogenesis, electroweak baryogenesis, and leptogenesis.
In this chapter, we take the electroweak baryogenesis and leptogenesis as the representative mechanisms to explain the matter-antimatter asymmetry of the universe.

\section{Electroweak baryogenesis}\label{firstparticles:sec3}

\begin{figure}[b]
\centering
\includegraphics[width=0.88\textwidth]{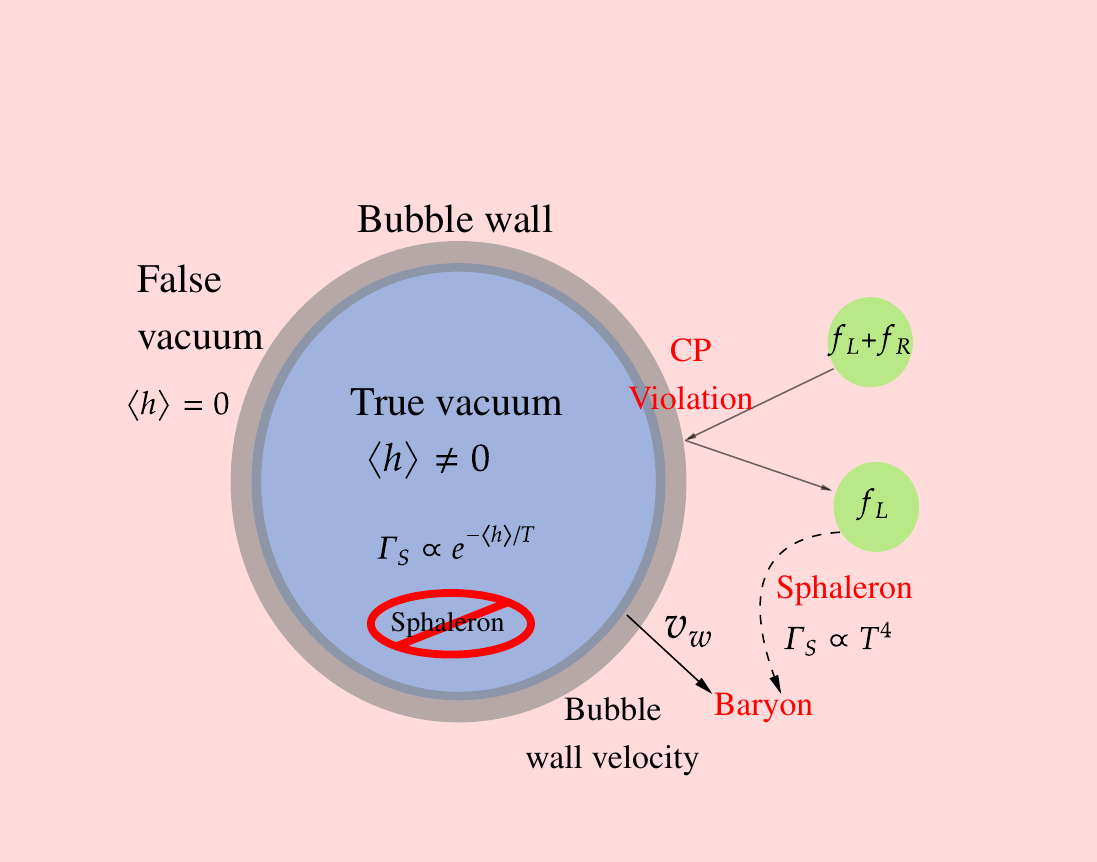}
\caption{ Schematic process of the electroweak baryogenesis. $f_L$ and $f_R$ are the left-handed and right-handed fermions respectively, and $v_w$ is the bubble wall velocity.}
\label{fig:ewbg}
\end{figure}

Electroweak baryogenesis became popular after the discovery of the Higgs boson at the Large Hadron Collider (LHC) and gravitational wave at The Laser Interferometer Gravitational-Wave Observatory (LIGO).
By extension of the Higgs sector, three Sakharov conditions can be satisfied, and the matter-antimatter asymmetry of the universe can be explained naturally.
The schematic process for the matter-antimatter asymmetry production in the electroweak baryogenesis is shown in Fig. \ref{fig:ewbg}.

Firstly, a strong electroweak first-order phase transition (FOPT) occurs when the temperature of the universe reduces to about 100 GeV with the expansion of the universe.
Bubbles are nucleated in this transition process.
Inside the bubbles, it is the true vacuum (broken phase, which means the vacuum expectation value of Higgs is nonzero, $\langle h \rangle \neq 0$), while outside the bubbles, it is the false vacuum (symmetry phase, $\langle h \rangle = 0$).
This FOPT naturally satisfies the third Sakharov condition for baryogenesis, namely, the departure from thermal equilibrium.

Secondly, the sphaleron process in the false vacuum violates the baryon number. This process satisfies the first Sakharov condition.
As can be seen in Fig. \ref{fig:ewbg}, with the CP violating scattering in front of the bubble walls, the net baryons are produced in the false vacuum. The scattering process needs C and CP violation to satisfy the 
 second Sakharov condition.
Inside the bubbles, there is a true vacuum, and the sphaleron process is greatly suppressed. So, no net baryons are produced in the bubbles. 

Lastly, as the bubbles expand, the true vacuum occupies the whole universe and absorbs the net baryons.

\subsection{Baryon number violation}
\begin{figure}[b]
\centering
\includegraphics[width=0.65\textwidth]{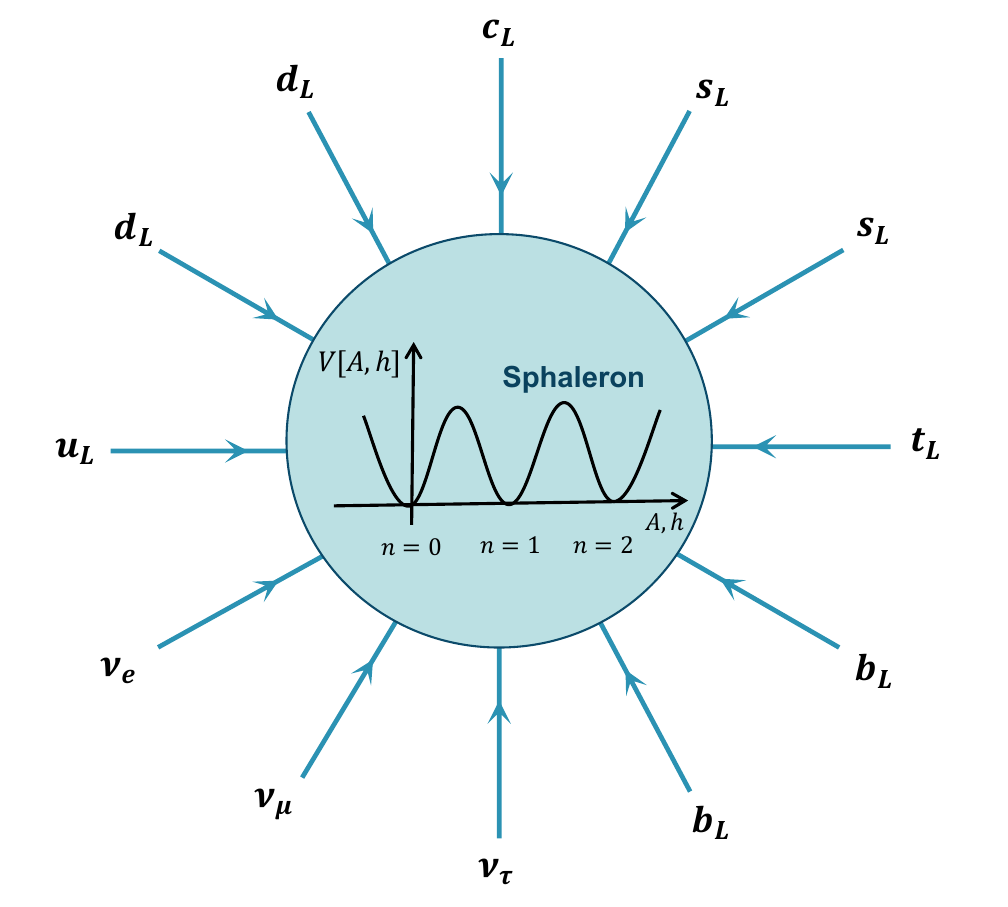}
\caption{Baryon number violation from the sphaleron process in the SM of particle physics. }
\label{fig:spha}
\end{figure}

By the end of the 1980s, extensive research had realized that the SM of particle physics might partially satisfy Sakharov's three conditions. 
In the SM, the classical Lagrangian possesses symmetries conserving both baryon and lepton numbers. However, at the quantum level, anomalous effects and the peculiarities of non-Abelian gauge field vacua break baryon and lepton numbers as shown in Fig. \ref{fig:spha}. 
This baryon number violation (sphaleron process) originates from the nonperturbative physics connected with the vacuum structure of chiral gauge theories.
More precisely, the electroweak vacuum is periodic, with vacua related to “large” gauge transformations.
However, the coupling to fermions is anomalous, which leads to baryon number violations when moving from one vacuum to another
\begin{equation}
	\partial_\mu j_B^\mu=\partial_\mu j_L^\mu=N_f\left(\frac{g^2}{32 \pi^2} W_{\mu \nu}^a \tilde{W}^{a \mu \nu}-\frac{g^{\prime 2}}{32 \pi^2} F_{\mu \nu} \tilde{F}^{\mu \nu}\right)
\end{equation}
where $W_{\mu \nu}^a$ and $F_{\mu \nu}$ are the field-strength tensors of the $SU(2)$ and $U(1)_Y$ gauge fields,  respectively. $g$ and $g'$ are the respective gauge coupling constants of the $SU(2)$ and $U(1)_Y$ gauge fields. $N_f$ denotes the number of fermion generations, and
$$
\tilde{F}^{\mu \nu}=\frac{1}{2} \epsilon^{\mu \nu \rho \sigma} F_{\rho \sigma}
$$
is the dual of $F^{\mu \nu}$. The expression for $\tilde{W}^{a \mu \nu}$ is analogous.
In fact, one can use this expression to compute the total change in baryon number for any such transition.
\begin{equation}\label{BminusL}
	\Delta B(t) = \Delta L(t)=\Delta N_{C S} \equiv N_f\left[N_{C S}\left(t\right)-N_{C S}(0)\right]
\end{equation}
where the Chern-Simons number is given by
\begin{equation}
	N_{C S}(t)=\frac{g_2^3}{96 \pi^2} \int d^3 x \epsilon^{i j k} \epsilon_{a b c} W_i^a W_j^b W_k^c(t)\,\,.
\end{equation}
The Chern-Simons number changes by one between adjacent vacua.
An important factor determining the rate of this reaction is the sphaleron process. The sphaleron is a static solution to the field equations in the electroweak model, representing the peak of the energy barrier between two distinct topological vacuum states. In the SM, the sphaleron process leads to baryon number violation and satisfies Sakharov's first condition (\cite{Kuzmin:1985mm}), as shown in Fig. \ref{fig:spha}.

Luckily, these effects are negligible at low temperatures, so protons remain stable.
Because at zero temperature, this is
a tunneling event, with a rate governed by the smallest
possible barrier between vacua, namely, the sphaleron energy
\begin{equation}
8~\mathrm{TeV}<E_{S}<14~\mathrm{TeV}
\end{equation}
At zero temperature, the baryon number violation (sphaleron) rate is negligible
\begin{equation}
	\Gamma_S(T=0) \sim \exp \left(-2 S_E\right) \sim 10^{-170}\,\,.
\end{equation}

However, when the temperature exceeds the electroweak scale (roughly corresponding to $T > \mathcal{O}(100)$ GeV), the processes breaking the baryon number will be in thermal equilibrium. 
At high temperatures, the sphaleron rate increases
\begin{equation}
	\Gamma_{S}(T)=\mu\left(\frac{M_W}{\alpha_W T}\right)^3 M_W^4 \exp \left(-\frac{E_{s p h}(T)}{T}\right)\,\,.
\end{equation}

In the false vacuum, as shown in Fig. \ref{fig:ewbg}, the sphaleron rate becomes large at high temperature
\begin{equation}
	\Gamma_S(T)=\kappa^{\prime} \alpha_W\left(\alpha_W T\right)^4 \quad \kappa^{\prime} \sim 30\,\,.
\end{equation}
However, in the true vacuum, the sphaleron rate is suppressed if the vacuum expectation value of Higgs over the temperature is larger than 1.
\begin{equation}
\Gamma_S \propto e^{-\langle h\rangle / T}\,\,,
\end{equation}
where we usually require 
\begin{equation}
\frac{\langle h\rangle}{T} >1\,\,.
\end{equation}
This is the so-called wash-out condition.

\subsection{Departure from thermal equilibrium: electroweak FOPT}

 Sakharov's third condition, a departure from thermal equilibrium, can be achieved through a strong electroweak FOPT. As shown in Fig. \ref{fig:ewbg}, initially, the electroweak symmetry $SU(2)_L\times U(1)_Y$ is unbroken, resulting in a net baryon number of zero. When the universe cools to approximately below 100 GeV, the electroweak phase transition occurs, generating matter-antimatter asymmetry. A successful asymmetry requires a strong  FOPT. During this transition, bubbles of broken phase, generated due to symmetry breaking in the symmetric phase plasma, expand, collide, and merge until the broken phase is completely established. Baryons are generated near the expanding bubble walls in the electroweak baryogenesis mechanism. Additionally, to generate a strong electroweak FOPT, the mass of the Higgs boson in the SM should be less than $\mathcal{O} (75)$ GeV~(\cite{Kajantie:1996mn}). So, to achieve strong electroweak FOPT, we often have to introduce new physics beyond the SM.

\begin{figure}[t]
\centering
\includegraphics[width=.8\textwidth]{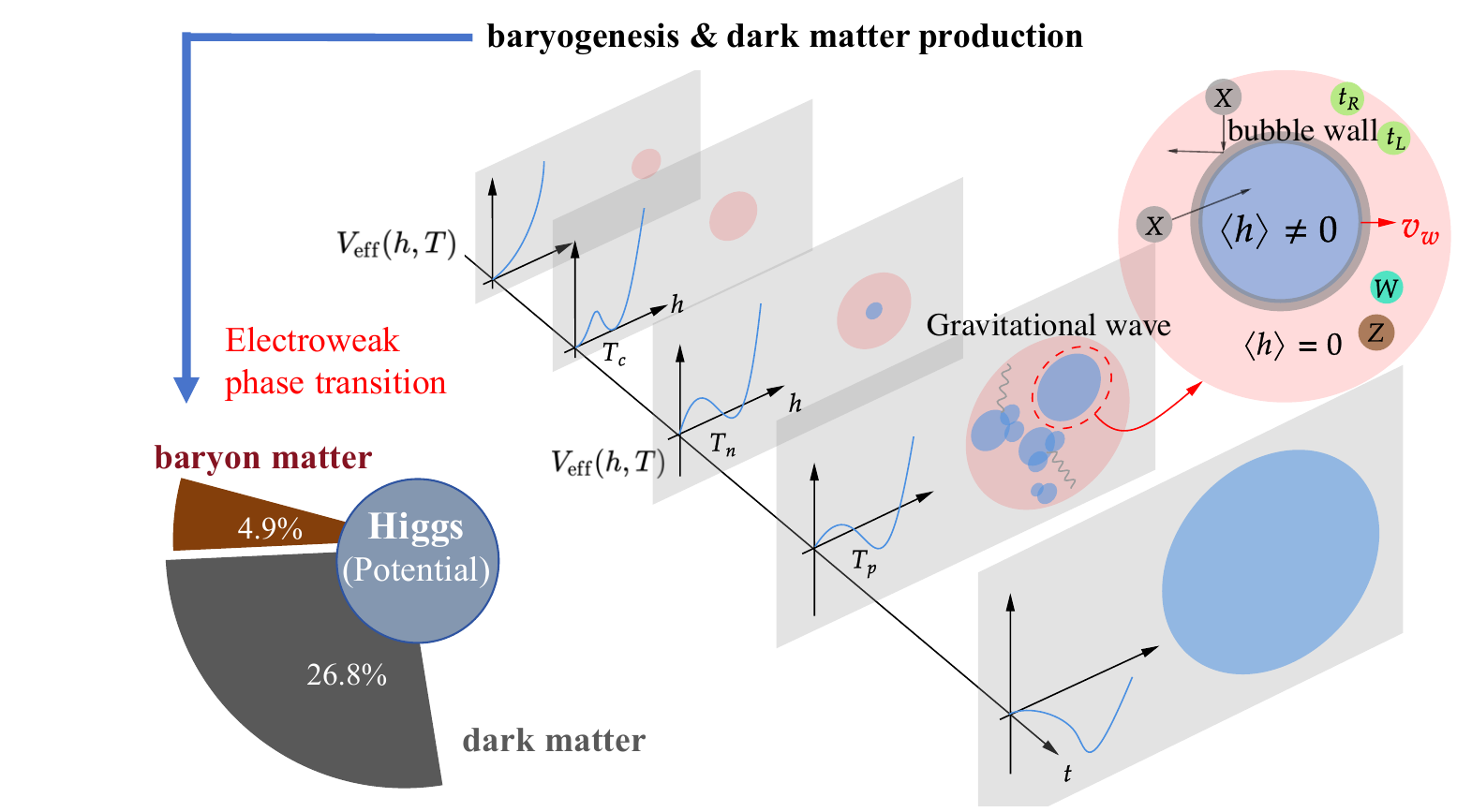}
\caption{The schematic process of the electroweak phase transition in the early universe. Matter-antimatter asymmetry and dark matter may be produced in this period.}
\label{fig:ewptgw}
\end{figure}

The electroweak FOPT in the early universe is shown in Fig.~\ref{fig:ewptgw}. As the temperature drops, a new stable state with lower potential energy appears, and the universe transits from the metastable state to a stable state. Bubbles nucleate, expand and collide with each other to produce gravitational wave. The bubble nucleation rate is determined by the free energy density of the early universe, which is equivalent to the finite temperature effective potential in the framework of thermal quantum field theory.
At one-loop level, the effective potential can be calculated as
\begin{equation}
V_{\mathrm{eff}}(h,T)=\sum_i n_i\left[\int \frac{\mathrm{d}^D p}{(2 \pi)^D} \ln \left(p^2+m_i^2(h)\right)+J_{\mathrm{B}, \mathrm{F}}\left(\frac{m_i^2(h)}{T^2}\right)\right]\,\,.
\end{equation}
where $n_i$ is the degree of freedom of particle $i$, $m_i$ is its mass and
\begin{equation}
    J_{\mathrm{B}, \mathrm{F}} (y)= \int dx x^2 \mathrm{ln}(1\mp e^{-\sqrt{x^2+y}})\,\,.
\end{equation}
Given the effective potential, we can calculate the action 
\begin{equation}
S(T) = \mathrm{min}[S_4(T),S_3(T)/T]\,\,,
\end{equation}
where
\begin{equation}
S_4 = 2\pi^2\int_{0}^{\infty} d\tilde{r} \tilde{r}^3 \left[\frac{1}{2}\left(\frac{d h}{d \tilde{r}}\right)^2+V_{\mathrm{eff}}(h, T)\right], \quad S_3 =4\pi \int dr r^{2} \left[\frac{1}{2}\left(\frac{d h}{d r}\right)^2+V_{\mathrm{eff}}(h, T)\right]
\end{equation}
with $\tilde{r} = \sqrt{(it)^2+r^2}$ where $t$ is the real time.
In both cases,
$h(r)$ is the $O(D=3,4)$-symmetric bounce solution and is determined by
\begin{equation}\label{bouncesolution}
    \frac{d^2 h }{dr^2} + \frac{D-1}{r}\frac{dh}{dr} = \frac{\partial V_{\mathrm{eff}}}{\partial h}\,\,.
\end{equation}
The equation \eqref{bouncesolution} can be solved by imposing the following boundary conditions:
\begin{equation}
\left.\frac{dh}{dr}\right|_{r=0} = 0, \quad h(r\rightarrow \infty) = 0\,\,.
\end{equation}

Finally, the nucleation rate can be obtained as
\begin{equation}
\Gamma=\Gamma_0 e^{-S(T)}\,\,.
\end{equation}
As shown in Fig.~\ref{fig:ewptgw}, after bubbles are nucleated, the bubbles expand, collide (\cite{Hogan:1983ixn,Witten:1984rs}), triggers turbulence (\cite{Kamionkowski:1993fg}) and sound waves. These processes can produce the so-called phase transition gravitational wave, which can help to explore these particle production processes between inflation and BBN.
The phase transition dynamics are essential to the baryogenesis and the corresponding gravitational wave signals.
The phase transition dynamics are described by the following phase transition parameters.
\begin{itemize}
\item[$\circ$]{The phase transition characteristic temperatures (critical temperature $T_c$, nucleation temperature $T_n$, and percolation temperature $T_p$) are defined by}
\begin{equation}
    V_{\mathrm{eff}}(0,T_c) =  V_{\mathrm{eff}}(v(T_c),T_c), \quad \frac{\Gamma(T_n)}{H_e(T_n)^4}=1, \quad \text{and} \quad  I(T_p) \equiv \frac{4\pi v_w^3}{3}\int_{T_p}^{T_c} \frac{dT'\Gamma(T')}{H_e(T')T'^4}\left( \int_{T_p}^{T'}\frac{d\tilde T}{H_e(\tilde T)}\right)^3 = 0.34\,\,,
\end{equation}
respectively. $v(T)$ is the vacuum expectation value and $H_e(T)$ is the Hubble expansion rate and is defined by
\begin{equation}
	H_e^2(T)=\frac{8 \pi}{3 M_{\mathrm{pl}}^2}\left(\frac{\pi^2}{30} g_* T^4+\Delta V_{\mathrm{eff}}(T)\right) \,\,.
\end{equation}
where $M_{\mathrm{pl}}=1.22 \times 10^{19} \,\,\mathrm{GeV}$  is the Planck mass, $g_*$ is the number of relativistic degrees of freedom. $\Delta V_{\mathrm{eff}}(T) = V_{\mathrm{eff}}(0,T) -V_{\mathrm{eff}}(v(T),T)$ is the potential difference between the false and true vacuum.

\item[$\circ$]{The phase transition strength $\alpha$ is defined by  
the ratio of the vacuum to the radiation energy density during the FOPT,}
\begin{equation}
    \alpha = \frac{\Delta V_{\mathrm{eff}} - T\frac{\partial \Delta V_{\mathrm{eff}}}{\partial T}}{\rho_R} \quad \text{with} \quad \rho_R = \frac{\pi^2}{30}g_* T^4\,\,.
\end{equation}
\item[$\circ$]{The phase transition duration $\beta$}
\begin{equation}
    \beta = H(T)\left.T\frac{d}{dT}\left(\frac{S_3}{T}\right)\right|_{T=T_n}
\end{equation}
and the mean bubble separation can be approximately estimated by $R_* = \frac{(8\pi)^{1/3}}{\beta}v_w$.

\item[$\circ$]{The bubble wall velocity $v_w$}

The bubble wall velocity is the most important phase transition parameter and is the most difficult to calculate.
When the bubbles expand in the Universe, they interact with all the particle content in a specific model. For example, during the electroweak phase transition, the background Higgs field interacts with Higgs bosons, quarks, leptons, and gauge fields. In the planar approximation, the equations of motion (EoM) of the background Higgs field in the fluid’s reference
frame reads (\cite{Moore:1995si})
\begin{equation}\label{eom3}
	(1-v_w^2)h'' + \frac{\partial  V_{\rm eff}(h,T)}{\partial h} +  \underbrace{\sum_i \frac{dm_i^2}{d h}\int\frac{d^3p}{(2\pi)^32E_i}\delta f_i(x,p)}_{\text{friction term}} = 0 \,\,.
\end{equation}
where the prime denotes the derivative with respect to $z$. Note that we have expanded the distribution function as $f=f_0+\delta f$ where $f_0$ represents the distribution in equilibrium. It can be seen that the out-of-equilibrium perturbations $\delta f$ on the particle distributions contribute to the friction term.

In order to extract the bubble wall velocity from the EoM of the background field, it is important to precisely calculate the friction term, or equivalently, the particles' perturbations $\delta f_i$. The precise calculation relies on the parametrization. In the 
traditional parametrization~(\cite{Moore:1995si,Wang:2020zlf,Jiang:2022btc}), the perturbations are parametrized by the chemical potential, temperature and velocity,
\begin{equation}\label{delta}
	f = \frac{1}{e^{(E+\delta)/T} \pm 1}, \quad \text{with} \quad \delta = -\mu - \mu_{bg}  - \frac{E}{T}(\delta T + \delta T_{bg}) - p_z(\delta v +\delta v_{bg})\,\,,
\end{equation}
where $\mu_{bg}$, $\delta T_{bg}$ and $\delta v_{bg}$ are quantities that refer to the background fluid.
However, this parametrization meets singularities at the sound velocity. In Ref.~(\cite{Laurent:2020gpg}), the authors propose a modified parameterization
\begin{equation}
    f = \frac{1}{e^{(E+\delta)/T} \pm 1} +\delta f_v, \quad \text{with} \quad \delta = -\mu - \mu_{bg}  - \frac{E}{T}(\delta T + \delta T_{bg})\,\,.
\end{equation}
Some other new methods have been proposed in recent years. The Ref.~(\cite{Dorsch:2021nje}) has pushed the formula \eqref{delta} to higher orders. In Refs.~(\cite{Laurent:2022jrs,DeCurtis:2022hlx}), the authors calculated the bubble wall velocity by first principles. Refs.~(\cite{Ai:2021kak,Ai:2023see}) propose that the bubble wall in local equilibrium also receives backreactions from hydrodynamic effects. After the parametrization, the main task is to solve the Boltzmann equations of the perturbations
\begin{equation}
    \left(\frac{p_z}{E}\partial_z - \frac{(m(z)^2)'}{2E}\partial_{p_z} \right)\delta f + \mathcal{C}\left[\delta f\right] = \frac{(m(z)^2)'}{2E}\partial_{p_z} f_0\,\,.
\end{equation}
After solving the Boltzmann equations and substituting the results of $\delta f$ into Eq.~\eqref{eom3}, we can solve the Eq.~\eqref{eom3} to extract the bubble wall velocity.

Besides, it is also important to investigate the run-away behavior of the bubble wall, which determines the shape of phase transition gravitational wave spectra, dark matter~(\cite{Azatov:2021ifm}), and baryogenesis~(\cite{Huang:2022vkf}). When the bubble wall velocity is relativistic, the behavior of bubble wall velocity is determined by the potential difference $\Delta V_{\mathrm{eff}}(T) $ and the leading-order (LO) friction
\begin{equation}
    \mathcal{P}_{LO} \simeq \sum_i g_i c_i \frac{\Delta m_i^2}{24} T^2\,\,,
\end{equation}
where $g_i$ is the degree of freedom and $c_i=1(1/2)$ for bosons (fermions). $\Delta m_i^2$ is the squared mass difference between the true and false vacuum. It can be seen that the LO friction is independent of the bubble wall velocity so that the run-away is possible. However, in the theories where the gauge bosons receive mass during the phase transition,  there would be additional next-to-leading-order (NLO) contributions. Ref.~(\cite{Hoche:2020ysm}) shows $\mathcal{P}_{NLO} \propto \gamma_w^2$ and (\cite{Gouttenoire:2021kjv}) gives $\mathcal{P}_{NLO} \propto \gamma_w$ where $\gamma_w = 1/\sqrt{1-v_w^2}$ is the Lorentz factor of the bubble wall. In both cases, the terminal steady bubble wall velocity can be extracted from the condition 
\begin{equation}
    \Delta V_{\mathrm{eff}} = \mathcal{P}_{LO} + \mathcal{P}_{NLO}\,\,.
\end{equation}

\item[$\circ$]{The energy budget}

The efficiency parameter $\kappa_v$ is defined by the ratio of the bulk kinetic energy of the fluid to the vacuum energy. It is an essential parameter for the calculation of gravitational waves from a FOPT. The calculation of the efficiency factor requires solving the self-similar hydrodynamic equations for the velocity profile $v$ and enthalpy profile $w$ of the fluid. The equations can be written in terms of $\xi = r/t$ where $r$ is
the distance from the bubble center and $t$ is the time since nucleation. The differential equations for $v(\xi)$ and $w(\xi)$ read~(\cite{Espinosa:2010hh,Wang:2020jrd})
\begin{equation}
    \gamma^2 (1-v\xi)\left[\left(\frac{\mu}{c_s} \right)^2-1 \right]\frac{dv}{d\xi} = \frac{2v}{\xi}\,\,, \quad \frac{1}{w}\frac{dw}{dv}=\frac{4\gamma^2\mu}{3c_s^2}\,\,,
\end{equation}
where $\mu(\xi,v)= (\xi-v)/(1-v\xi)$ and $c_s$ is the sound velocity. Generally, one uses the bag model where the equation of state can be written as $p=\frac{aT^4}{3}-\epsilon$ and $e = aT^4 + \epsilon$ with $p$ and $e$ are  the pressure and energy density respectively. $a=\frac{\pi^2 g_*}{30}$ and $\epsilon$ is the bag constant describing the difference in energy density and pressure across the bubble wall. In the bag model $c_s=1/\sqrt{3}$ and it is meaningful to consider the energy budget beyond bag model~(\cite{Giese:2020znk,Wang:2020nzm}).
The efficiency parameter is then given by
\begin{equation}
    \kappa_v = \frac{3}{\Delta\epsilon v_w^3} \int w(\xi) v^2 \gamma^2 \xi^2 d\xi\,\,.
\end{equation}
\end{itemize}
where $\Delta\epsilon$ corresponds to the energy density difference between the
symmetric phase and broken phase.
The model-dependent analysis method for the energy budget of the FOPT can be seen in Ref.~(\cite{Wang:2023jto}).

In 2012, the Large Hadron Collider (LHC) discovered the Higgs boson~(\cite{ATLAS:2012yve,CMS:2012qbp}) with a mass of 125 GeV (awarded the Nobel Prize in Physics in 2013), directly refuting the possibility of achieving electroweak baryogenesis because the 125 GeV Higgs cannot realize the strong FOPT. 
The SM of particle physics cannot explain the asymmetry between matter and antimatter in the universe. 
To achieve a successful electroweak baryogenesis mechanism, we must extend the SM of particle physics, particularly in the Higgs sector. Some common models in this regard include effective field theories, multi-Higgs models, left-right symmetric electroweak unification models, etc. Compared to the SM, these extended models offer new sources of CP violation and a richer Higgs particle spectrum. 
 
\subsection{C and CP violation}

In the SM, as evidenced by the Cabibbo-Kobayashi-Maskawa (CKM) matrix, both C and CP symmetries are broken. 
However, quantitative calculations show that the amount of CP violation provided by the CKM matrix in the SM is insufficient to generate the baryon asymmetry of the universe.
Extra CP violating sources are needed.

There is a strong tension between the electric dipole moment (EDM) experiments and the traditional
electroweak baryogenesis to explain the observed matter-antimatter asymmetry.
In traditional electroweak baryogenesis models, one needs large enough
CP-violating source for successful electroweak baryogenesis.
However, only a pretty small CP violation is allowed to avoid strong EDM constraints.
The recent EDM measurements data give (\cite{Roussy:2022cmp})]
\begin{equation}
	\left|d_e\right|<4.1 \times 10^{-30}\quad e\cdot\mathrm{~cm}
\end{equation}

How can we alleviate this tension for successful baryogenesis?
One natural idea is dynamical CP violation for baryogenesis, where the CP violation source evolves 
with the evolution of the universe. A concrete approach to realizing dynamical CP violation is through the two-step cosmic phase transition in the effective field theory~(\cite{Huang:2018aja}).
Dynamical CP violation might also appear in the renormalizable model Complex 2HDM~(\cite{Wang:2019pet}).
Generally, one needs to solve the transport equations to quantify how the CP violation source can be transformed into the final baryon asymmetry.

\subsection{Transport equations}
 Precise calculation of electroweak baryogenesis requires calculating the CP violation source by solving the transport equations. Particles will feel the mass variation when they cross the bubble wall
\begin{equation}
\mathcal{M}=m(z) e^{i \theta(z)}\,\,.
\end{equation}
According to energy conservation and Newton's second law of motion, the mass variation is equivalent to the external force from the bubble wall. The transport equations can be obtained through the Wentzel–Kramers–Brillouin (WKB) approximation (\cite{Joyce:1994fu,Cline:1997vk,Cline:2000kb}) or Kadanoff-Baym approach~(\cite{Kainulainen:2001cn}). In the planar approximation,  the evolution of the distribution of particles with one flavor follows:
\begin{equation}\label{transport}
\left(v_g \partial_z+F \partial_{p_z}\right) f=\mathcal{C}[f]\,\,,
\end{equation}
where
\begin{equation}
\begin{gathered}
v_g=\frac{p_z}{E}+s s_{k_0} \frac{m^2 \theta^{\prime}}{2 E^2 E_z}\,\,, \\
F=-\frac{\left(m^2\right)^{\prime}}{2 E}+s s_{k_0}\left(\frac{\left(m^2 \theta^{\prime}\right)^{\prime}}{2 E E_z}-\frac{m^2\left(m^2\right)^{\prime} \theta^{\prime}}{4 E^3 E_z}\right)\,\,,
\end{gathered}
\end{equation}
where $s_{k_0}=\pm 1$ for particles and antiparticles respectively. $s= \pm 1$ for the eigenstates of the spin $s$ in $z$-direction in the frame where the momentum parallel to the wall vanishes.
Note that $E \equiv \sqrt{\mathbf{p}^2+m^2}$ and $E_z \equiv \sqrt{p_z^2+m^2}$.

The equation \eqref{transport} is the central point for electroweak baryogenesis. The kinetic term $v_g$ in combination with the
collision terms on the right side, dictates how the particle densities diffuse away from the bubble wall and also determines how the asymmetries are communicated
to the other particle species. The forces on the left-hand side $F$ encode how the plasma is driven out-of-equilibrium and how CP violation manifests
itself in the particle distribution.

 Historically, in order to derive the WKB fluid equations, the starting point is the Boltzmann equation for the asymmetries between the particle and antiparticle distribution functions, $\delta f_i(x, p)$, which is subject to an external force from the bubble wall. Splitting $F$ into its $C P$-conserving and $C P$-violating parts $F_0+\delta F$,
\begin{equation}\label{perturb}
\left(\frac{p_z}{E} \partial_z+F_0 \partial_{p_z}\right) \delta f_i=\mathcal{C}\left[\delta f_j\right]-\delta F \partial_{p_z} f_{0} \,\,.
\end{equation}
Here $\mathcal{C}$ is the collision term, linearized in the asymmetries $\delta f_j$. $f_0$ is the unperturbed distribution function in equilibrium. The last term on the right side leads to the $C P$-violating sources.

The next step is to approximate the full Boltzmann equation by a set of fluid equations for the chemical potential $\mu_i$, which corresponds to the particle number density, and one higher moment  $u_i$, which corresponds to the velocity perturbation. This can be done by integrating equation \eqref{perturb} over spatial momenta, weighted by 1 and $p_z / E$, respectively. The CP-violating source term, which is the phase space average of $\vec{p} \cdot \delta \vec{F} / E$, appears principally in the fluid equation for $u_i$. This is because $\vec{\nabla}_p \delta f_i$ nearly averages to zero unless it is weighted by $p_z$. After all, the generic form of the WKB fluid equations is~(\cite{Cline:2021dkf})
\begin{equation}\label{matrix}
A_i\binom{\mu_i}{u_i}^{\prime}+\left(m_i^2\right)^{\prime} B_i\binom{\mu_i}{u_i}-\mathcal{C}_i^{W K B}=\binom{O\left(v_w S_{W K B, i}\right)}{S_{W K B, i}}\,\,,
\end{equation}
where $A_i$ and $B_i$ are dimensionless matrices that depend upon $m_i^2(z) / T^2$ and $v_w$. The chemical potential
for left-handed baryon number is
\begin{equation}
\mu_{B_{\mathrm{L}}}=\frac{1}{2}\left(1+4 D_0^t\right) \mu_{t_{\mathrm{L}}}+\frac{1}{2}\left(1+4 D_0^b\right) \mu_{b_{\mathrm{L}}}+2 D_0^t \mu_{t_{\mathrm{R}}} 
\end{equation}
where $t_L$ is the left-handed top quark, $b_L$ is the left-handed bottom quark and $t_R$ is the right-handed top quark. $D_0^i$ is defined by
\begin{equation}
D_0^i = \frac{1}{N_1}\int d^3p f_0', \quad \text{with} \quad N_1 = -\frac{2}{3}\pi^3 T^2\gamma_w\,\,,
\end{equation}
where the prime for $f_0'$ denotes $d/d(\gamma_w E)$.
The baryon asymmetry can be calculated by integrating the sphaleron rate equation.
\begin{equation}
	\eta_B=\frac{405 \Gamma_{S}}{4 \pi^2 \gamma_w v_w g_* T} \int d z \mu_{B_L} f_{\mathrm{sph}} e^{-45 \Gamma_{S}|z| / 4 \gamma_w v_w}\,\,.
\end{equation}
The function $f_{\mathrm{sph}}(z)=\min \left(1,2.4 \frac{\Gamma_{S}}{T} e^{-40 h(z) / T}\right)$ is smoothly interpolation function of sphaleron rates between the broken and unbroken phases. $g_*$ is the number of degrees of freedom of the thermal bath.

There is another calculation method that is also commonly used —— the vacuum expectation value insertion approximation (VIA) method~(\cite{Riotto:1995hh,Riotto:1997vy}).
Unlike the WKB formalism, the VIA method uses second order fluid equations, with half dependent variables per particle species. The basic form of the fluid equations in the VIA approximation reads:
\begin{equation}
    D_i n_i^{\prime \prime}+v_w n_i^{\prime}-\mathcal{C}_i^{V I A}\left[n_j\right]=S_{V I A, i}\,\,,
\end{equation}
where $D_i$  is the diffusion constant, and the source term is given by~\cite{Riotto:1998zb}
\begin{equation}
S_{\mathrm{VIA}}(x) \sim \int d^4 y\left(m_x m_y^*-m_x^* m_y\right) \operatorname{Im} \operatorname{tr}\left[G_{x y}^{>} G_{y x}^{<}\right]\,\,,
\end{equation}
where $m_x\equiv m(x)$ and $G$ is the propagator in thermal field theory.

It can be shown that the WKB method can give the same form of fluid equations by combining equation \eqref{matrix} to cancel the $u_i$ and using the fact $n=\mu T^2/6$. Recently, in Refs.~(\cite{Cline:2020jre}) and (\cite{Cline:2021dkf}), James Cline and his collaborators made a comparison between the two methods in a specific model, they found that the predictions typically differed by factors of $10^1-10^2$. The main reason for the discrepancy is the different definitions of the source term.
Recently, it has been pointed out that the VIA source term
vanishes precisely when an accurate resummation of the one-particle irreducible self-energy is performed, ensuring all relevant quantum corrections are incorporated consistently into the propagator, leading to a fully renormalized and stable result~(\cite{Kainulainen:2021oqs,Postma:2022dbr}).

\subsection{Benchmark scenario of electroweak baryogenesis}
The most appealing aspect of electroweak baryogenesis is its close connection to the Higgs boson, whose properties can be tested in collider experiments and gravitational wave experiments. Generally, electroweak baryogenesis requires a strong FOPT, modifying the trilinear Higgs boson interaction vertices in the SM. This can be determined by measuring the invariant mass distribution of Higgs boson pairs produced in hadron colliders (unfortunately, the capabilities of the 14 TeV LHC are insufficient to measure this coupling parameter, but this prediction may be verified by the future Super Proton-Proton Collider with a capability of 100 TeV). Of course, more precise verification can be achieved by measuring the cross-section of Higgs boson production in association with Z bosons at electron colliders, such as the Circular Electron-Positron Collider and the International Linear Collider. Additionally, during the phase transition of early-universe electroweak baryogenesis, gravitational waves are produced by collisions between bubbles and the turbulence between bubbles and the plasma. Future space-based gravitational wave interferometry experiments (such as LISA, TianQin, and Taiji) can detect this signal with observable sensitivity. Experiments on gravitational waves in space and collider experiments on Earth complement each other, helping us understand the origin of matter, the loss of antimatter, and the nature of the Higgs boson.

There are various Higgs extended models to realize the electroweak baryogenesis.
We show representative examples.
The first example is the dimension-6 Higgs model as below.
\begin{equation}
	\delta \mathcal{L}=-x_u^{i j} \frac{H^{\dagger} H}{\Lambda^2} \bar{Q}_{L i} \tilde{H} u_{R j}+\text { H.c. }-\frac{\kappa}{\Lambda^2}\left(H^{\dagger} H\right)^3\,\,.
\end{equation}
This effective Lagrangian can be obtained from many new physics models from the perspective 
of the SM effective field theory~(\cite{Huang:2015izx,Huang:2016odd,Huang:2015izx,Huang:2016odd,Cao:2017oez}). The first term represents the CP-violating interaction between the Higgs boson and the fermions.
The last term can trigger a strong FOPT.

Another example is the so-called two-step phase transition model~(\cite{Huang:2018aja,Espinosa:2011eu,Cline:2012hg}).
The Lagrangian density for this model is given by
\begin{equation}
	\mathcal{L}= \mathcal{L}_{\mathrm{SM}}-y_t \frac{\eta}{\Lambda} S \bar{Q}_L \tilde{H} t_R+\text { H.c }+\frac{1}{2} \partial_\mu \mathrm{S} \partial^\mu \mathrm{S}+\frac{1}{2} \mu^2 \mathrm{~S}^2-\frac{1}{4} \lambda \mathrm{S}^4-\frac{1}{2} \kappa \mathrm{S}^2\left(H^{\dagger} H\right)\,\,.
\end{equation}
The second term can provides large CP-violating source, and the
last three terms can induce two-step phase transition.









\section{Leptogenesis}\label{firstparticles:sec4}

In recent years, due to the advancements in neutrino oscillation physics, mechanisms for generating lepton number asymmetry have garnered considerable attention. In the SM, the baryon and lepton numbers are violated by sphaleron processes, as illustrated in Fig. \ref{fig:spha}, but their difference is conserved, which can be seen from Eq.~\eqref{BminusL}. This links changes in baryon numbers to changes in lepton numbers, as baryon number asymmetry can be converted from lepton number asymmetry through sphaleron processes. Generally, mechanisms for generating lepton number asymmetry require processes that violate lepton number, CP and CPT violations in the lepton sector and attain a non-equilibrium state. These conditions can be realized in models describing massive neutrinos. For example, in the simple seesaw model, neutrinos are of the Majorana type, breaking lepton number symmetry, and the decoupling of heavy right-handed neutrinos provides the non-equilibrium condition. However, no experimental evidence indicates that neutrinos must be of the Majorana type. In other words, neutrinos could potentially be Dirac fermions, like other charged fermions. In such a scenario, mechanisms for generating lepton number asymmetry can still be realized. This is because sphaleron processes only directly affect left-handed fermions, and the left-handed neutrinos and right-handed neutrinos only reach thermal equilibrium at very low temperatures when sphaleron processes are no longer effective. Consequently, if there is a lepton number for a left-handed lepton and an opposite lepton number for a right-handed neutrino, even though the total lepton number is strictly zero, there is no lepton number asymmetry. Sphaleron processes can then partially convert the lepton number of left-handed leptons into baryon number, thus explaining baryon number asymmetry. This lepton number-conserving mechanism for generating lepton number asymmetry can be realized in some Dirac neutrino models. Typically, the scale of mechanisms for generating lepton number asymmetry is very high, making it difficult to test. In some specific models, mechanisms for generating lepton number asymmetry can predict CP violation and the mass ordering of neutrinos, which can be tested in neutrino oscillation experiments and neutrinoless double beta decay experiments.
We present more details below.

\subsection{lepton asymmetry from heavy neutrino decay}

\begin{figure}[t]
	\centering
	\includegraphics[width=0.88\textwidth]{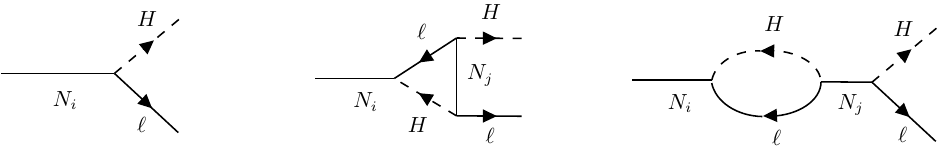}
	\caption{ Tree level and one-loop contributions to the heavy Majorana neutrino decay.}
	\label{fig:decay}
\end{figure}

To describe traditional thermal leptogenesis, we can extend the SM by 3 right-handed neutrino fields:
\begin{equation}
\mathcal{L}=\mathcal{L}_{S M}+i \overline{N_{R i}} \gamma_\mu \partial^\mu N_{R i}-y_i \overline{\ell_L}\tilde{H} N_{R i} -\frac{1}{2} \overline{\left(N_{R i}\right)^c} M_{i} N_{R i}+\text { H.c. },
\end{equation}
where $i=1,2,3$ and flavor indices have been suppressed. $H, \ell_L$, and $N_R$ denote the SM $SU(2)_L$ Higgs-doublet, the $SU(2)_L$ lepton doublet and the right-handed $SU(2)_L \times$ $U(1)_Y$ singlet neutrino respectively. $y$ denotes the corresponding Yukawa couplings and $\tilde H = i\sigma_2 H^*$ is the charge conjugation of
the Higgs doublet. After spontaneous electroweak symmetry breaking, the neutrinos acquire the Dirac masses
\begin{equation}
m_D=y v_{EW},
\end{equation}
with $v_{EW} \simeq 246~\mathrm{GeV}$ being the Higgs vacuum expectation value. The model predicts three heavy right-handed Majorana neutrinos with masses $M_i$, which can explain the small masses of the light neutrinos in terms of the see-saw mechanism. The heavy Majorana neutrinos can decay into lepton-Higgs pairs via the Yukawa coupling terms:
\begin{equation}
    N_i \rightarrow \ell H, \quad N_i \rightarrow \bar{\ell} \bar{H}\,\,.
\end{equation}
The process must be CP asymmetric, and the CP asymmetry in the decay 
is caused by interference between the tree level and the one-loop diagrams. See Fig.~\ref{fig:decay}.

The CP asymmetry is defined by
\begin{equation}
\epsilon_i=\frac{\Gamma_{N_i \rightarrow \ell H}-\Gamma_{N_i \rightarrow \bar{\ell} \bar{H}}}{\Gamma_{N_i \rightarrow \ell H}+\Gamma_{N_i \rightarrow \bar{\ell} \bar{H}}}\,\,,
\end{equation}
where $\Gamma_{N_i \rightarrow \ell \phi}$ includes a sum over flavour indices. At one-loop order, we can write the decay
amplitudes as $\mathcal{M}_{N_i \rightarrow \ell H}=g_0 \mathcal{M}_0+g_1 \mathcal{M}_1$ and $\mathcal{M}_{N_i \rightarrow \bar{\ell} \bar{H}}=g_0^* \mathcal{M}_0+g_1^* \mathcal{M}_1$, where $g_0 \mathcal{M}_0$ and $g_1 \mathcal{M}_1$ are the tree-level and one-loop level contributions respectively. $g_0, g_1$ represent the products of all coupling constants in these diagrams. The CP-violating parameter then reads
\begin{equation}
\epsilon_i=\frac{\left|g_0 \mathcal{M}_0+g_1 \mathcal{M}_1\right|^2-\left|g_0^* \mathcal{M}_0+g_1^* \mathcal{M}_1\right|^2}{\left|g_0 \mathcal{M}_0+g_1 \mathcal{M}_1\right|^2+\left|g_0^* \mathcal{M}_0+g_1^* \mathcal{M}_1\right|^2} \simeq-2 \frac{\Im\left\{g_0^* g_1\right\} \Im\left\{\mathcal{M}_0^* \mathcal{M}_1\right\}}{\left|g_0\right|^2\left|\mathcal{M}_0\right|^2}\,\,.
\end{equation}

It is easy to see that $\epsilon_i$ has to be zero when we include only tree level contributions. $\Im $ represents the imaginary part.
Both $g_0^* g_1$ and $\mathcal{M}_0^* \mathcal{M}_1$ need to have non-vanishing imaginary parts in order to get nonzero $\epsilon_i$. The former implies we must have  $j\neq i$ in the loop graphs in Fig.~\ref{fig:decay}. This means that at least two heavy Majorana neutrinos need to be included.
We assume the lepton asymmetry is produced by the lightest heavy neutrino $N_1$, the asymmetry from the decay of $N_{2,3}$ is washed out by the process including $N_1$. Then, the CP asymmetry can be calculated from Fig.~\ref{fig:decay} :
\begin{equation}
\epsilon_1=\frac{3}{16 \pi} \sum_{j=2,3} \frac{\Im\left\{\left(y^{\dagger} y\right)_{1 j}^2\right\}}{\left(y^{\dagger} y\right)_{11}} f\left(\frac{M_j^2}{M_1^2}\right)\,\,,
\end{equation}
where the loop function reads
\begin{equation}
f(x)=\frac{2}{3} \sqrt{x}\left[\frac{2-x}{1-x}-(1+x) \ln \left(\frac{1+x}{x}\right)\right] \underset{x \rightarrow \infty}{\longrightarrow} -\frac{1}{\sqrt{x}} \,\,.
\end{equation}
When we assume $M_{1} \ll M_{2},M_{3}$, we have
\begin{equation}
\epsilon_1 \simeq -\frac{3}{16 \pi} \sum_{j=2,3} \frac{\Im\left\{y_{1\alpha}^*y_{j \beta} y_{1\beta}^* y_{j \alpha}\right\}}{\left(y_{1 \alpha} y_{1\alpha}^*\right)} \frac{M_1}{M_j} \,\,.
\end{equation}
Note that we have added the flavor indices here. During the evolution of lepton asymmetry in the universe, the produced asymmetry also suffered the washout effects from inverse decay and scatterings. Washout represents some reactions that can depletes the generated lepton asymmetry. The weak and strong washout regimes correspond, respectively, to the
limits $K \ll 1$ and $K \gg 1$ where the decay parameter $K$ is defined by 
\begin{equation}
K=\frac{\left.\Gamma_D\right|_{T=0}}{\left.H\right|_{T=M_1}}=\frac{\frac{1}{8 \pi}\left(y y^{\dagger}\right)_{11} M_1}{1.66 g_*^{1 / 2} \frac{M_1^2}{M_{Pl}}}=\frac{\left(y y^{\dagger}\right)_{11} \frac{v_{\mathrm{EW}}^2}{M_1}}{8 \pi \times 1.66 g_*^{1 / 2} \frac{v_{\mathrm{EW}}^2}{M_{Pl}}} \equiv \frac{\widetilde{m_1}}{m_*}\,\,,
\end{equation}
where we have defined an ``effective neutrino mass":
\begin{equation}
\widetilde{m_1} \equiv\left(y y^{\dagger}\right)_{11} \frac{v_{\mathrm{EW}}^2}{M_1} \,\,,
\end{equation}
and an ``equilibrium neutrino mass":
\begin{equation}
m_* \equiv 8 \pi \times 1.66 g_*^{1 / 2} \frac{v_{\mathrm{EW}}^2}{M_{\mathrm{pl}}} \simeq 10^{-3} \mathrm{eV}\,\,.
\end{equation}
The factor $K$ also evaluates to what extent the heavy neutrinos are out of equilibrium. For $K \gg 1$ the heavy neutrinos are in equilibrium and for 
$K \ll 1$ the heavy neutrinos are far out of equilibrium.

\subsection{Conversion of the lepton asymmetry into a baryon asymmetry}
After the production of lepton asymmetry, the sphaleron process will convert it into baryon asymmetry. The relationship is decided by the processes in equilibrium in the early universe. Firstly, the number asymmetry for each particle species $k$ can be related to the chemical potential as follows:
\begin{equation}
n_k-\bar{n}_k\simeq \frac{g T^3}{6}\left\{\begin{array}{cc}
\mu_k / T & \text { for fermions } \\
2 \mu_k / T & \text { for bosons }
\end{array}\right.
\end{equation}
at lowest order for small $\mu_k/T$.
Thus, the asymmetry between the number of baryons and antibaryons reads:
\begin{equation}
n_B-n_{\bar{B}} \simeq \frac{g T^3}{6} \sum_{k=1}^{N_f}\left(2 \mu_{Q_i}+\mu_{u_i}+\mu_{d_i}\right)\,\,,
\end{equation}
where $i$ is the generation of quarks and $N_f$ is the number of fermion generations. In SM the number of fermion generations is three such that $N_f=3$.
Similar form holds for the lepton asymmetry:
\begin{equation}
n_L-n_{\bar{L}} \simeq \frac{g T^3}{6} \sum_i^{N_f}\left(2 \mu_{L_i}+\mu_{e_i}\right)\,\,.
\end{equation} 
We define
\begin{equation}
B=\sum_{i=1}^{N_f}\left(2 \mu_{Q_i}+\mu_{u_i}+\mu_{d_i}\right), \quad L = \sum_i^{N_f}\left(2 \mu_{L_i}+\mu_{e_i}\right)\,\,.
\end{equation}
Although it seems these chemical potentials are  independent quantities, they fulfill the following relations when the particles are in chemical equilibrium in these processes:
\begin{itemize}
\item[$\circ$] The total hypercharge of the plasma must vanish:
\begin{equation}
\sum_i^{N_f}\left(\mu_{Q_i}+2 \mu_{u_i}-\mu_{d_i}-\mu_{e_i}+\frac{2}{N_f} \mu_{H}\right)=0 \,\,.
\end{equation}
\item[$\circ$] If the Yukawa interactions through the Higgs portal are in thermal equilibrium,
\begin{equation}
\begin{aligned}
\mu_{Q_i}-\mu_{H}-\mu_{d_j}=0, \quad
\mu_{Q_i}+\mu_{H}-\mu_{u_j}=0, \quad 
\mu_{L_i}-\mu_{H}-\mu_{e_j}=0 \,\,.
\end{aligned}
\end{equation}
\item[$\circ$] The 12-fermion interactions induced by sphalerons (which can be seen from Fig \ref{fig:spha}) lead to:
\begin{equation}
\sum_i\left(3 \mu_{Q_i}+\mu_{L_i}\right)=0 \,\,.
\end{equation}
\item[$\circ$] QCD instanton processes lead to interactions between left-handed and right-handed quarks, which are described by the operator $\prod_i\left(Q_{L_i} Q_{L_i} u_{R_i}^c d_{R_i}^c\right)$. If these processes  are in equilibrium, they lead to:
\begin{equation}
\sum_i\left(2 \mu_{Q_i}-\mu_{u_i}-\mu_{d_i}\right)=0 \,\,.
\end{equation}
\end{itemize}
If we have equilibrium among different generations, we can get $\mu_{L_i} \equiv \mu_L, \mu_{Q_i} \equiv \mu_Q$, etc. All chemical potentials can be written in terms of $\mu_L$ :
\begin{equation}
\begin{gathered}
\mu_e=\frac{2 N_f+3}{6 N_f+3} \mu_L, \quad \mu_u=\frac{2 N_f-1}{6 N_f+3} \mu_L, \quad \mu_d=\frac{6 N_f+1}{6 N_f+3} \mu_L,
\end{gathered}
\end{equation}
\begin{equation}
\mu_Q=-\frac{1}{3} \mu_L, \quad \mu_{H}=\frac{4 N_f}{6 N_f+3} \mu_L\,\,.
\end{equation}

Therefore, we have
\begin{equation}
B=-\frac{4}{3} N_f \mu_L, \quad L=\frac{11 N_f^2+9 N_f}{6 N_f+3} \mu_L\,\,,
\end{equation}
from which one can conclude that leptogenesis generates a $B$ asymmetry. Concretely,
\begin{equation}
B=c(B-L), \quad L=(c-1)(B-L) \,\,,
\end{equation}
with
\begin{equation}
c=\frac{8 N_f+4}{22 N_f+13} \,\,.
\end{equation}
In the SM $c= 28/79$, then we get
\begin{equation}
    B = -\frac{28}{51}L\,\,.
\end{equation}
This is the relation between the produced lepton asymmetry and the final baryon asymmetry.

\subsection{Developments of leptogenesis}

The CP asymmetry is bounded from above for a hierarchical spectrum of right-handed neutrinos, known as the Davidson-Ibarra bound~(\cite{Davidson:2002qv}):
\begin{equation}\label{DI}
|\epsilon_1| \lesssim \frac{3M_1(m_3-m_1)}{8\pi v_{\mathrm{EW}}^2}\approx 10^{-5} \times\left(\frac{M_1}{10^{11} \mathrm{GeV}}\right)\left(\frac{m_3-m_1}{0.05 \mathrm{eV}}\right)\,\,.
\end{equation}
By using the parametrization of the baryon to entropy ratio $Y_B=n_B/s$ and $ Y_{N_1}^{\mathrm{eq}}=n_{N_1}^{\mathrm{eq}}/s$ where $n_{N_1}^{\mathrm{eq}}$ is the equilibrium number density of the lightest right-handed neutrino, the yield of the generated baryon asymmetry reads
\begin{equation}
Y_B= Y_{N_1}^{\mathrm{eq}} \epsilon_{1} \kappa_{\text {sph }} \kappa_{\text {wash }}\,\,,
\end{equation}
where $\kappa_{\text {sph }} $ accounts for the fraction of the lepton
asymmetry converted to baryon asymmetry by sphalerons, and $\kappa_{\text {wash }} \lesssim 1$ accounts for suppression due to the washout processes. For $\kappa_{\text {wash }} \sim 10^{-2}-10^{-3}$, from equation \eqref{DI} we have roughly $M_1 \sim 10^{11}~\mathrm{GeV}$. So it is challenging to test this minimal thermal leptogenesis model. Some new mechanisms can realize leptogenesis at $\mathcal{O}(\mathrm{TeV})$. For example, if the masses of at least two generations of heavy neutrinos are nearly degenerate, the CP violating phase can
be resonantly enhanced~(\cite{Pilaftsis:2003gt}). Besides, the right-handed neutrinos could couple to left-handed neutrinos via a new dark Higgs doublet~(\cite{Clarke:2015hta}). Recently, it has been shown that the energy scale of thermal leptogenesis can be suppressed by including the B-L breaking phase transition dynamics, the FOPT  gravitational wave signal can help to detect this leptogenesis model~(\cite{Huang:2022vkf}).

\section{Conclusion}
In this chapter, we present a brief review of the mechanism of reheating and baryogenesis. The reheating process heats the cold, uniform, homogeneous matter of the initial period at the late stage of cosmic inflation into the ultra-hot various species of matter, in place at the onset of the BBN. Studying particle production in a concrete particle physics model and the evolution of cosmological perturbations during reheating are still active research directions.

The origin of the observed matter-antimatter asymmetry in the universe remains largely unknown. This is a significant issue in both particle physics and cosmology, and it closely correlates with these two fields. The two mechanisms for baryogenesis highlighted above (electroweak baryogenesis and leptogenesis) focus on explaining the origin of matter-antimatter asymmetry from the perspective of particle physics. Both mechanisms, especially the electroweak baryogenesis mechanism, require an extension of the SM of particle physics and have the potential to be tested in particle physics experiments and gravitational wave detectors.
To fully determine which mechanism the matter-antimatter asymmetry originates from, joint efforts from theoretical and experimental physicists are needed.

Thus, any extension of SM should be able to explain the reheating process, the baryogenesis, and dark matter production. This makes astronomy and cosmology another test of our fundamental particle models.

\begin{ack}[Acknowledgments]

FPH acknowledges support from the National Natural Science Foundation of China (NNSFC) under Grant No. 12475111.
FPH also extends gratitude to Siyu Jiang for his valuable contributions to enhancing the manuscript.
\end{ack}

\seealso{Inflation; Big Bang Nucleosynthesis; dark matter}

\bibliographystyle{Harvard}
\bibliography{reference}

\end{document}